\documentstyle[aas2pp4]{article}
\def\jnl@aj{AJ}

\received{... 1999}
\accepted{... 1999}

\journalid{00}{... 1999}
\articleid{00}{... 1999}

\lefthead{Caretta, Maia \& Willmer}
\righthead{An Evaluation of COSMOS and APM Catalogs}

\begin{document}

\title{A Quantitative Evaluation of the Galaxy Component of \\
COSMOS and APM Catalogs\footnote{Partly based on observations at 
Complejo Astronomico El Leoncito (CASLEO), operated under 
agreement between the Consejo Nacional de Investigaciones 
Cient\'\i ficas de la Rep\'ublica Argentina and the National 
Universities of La Plata, C\'ordoba and San Juan; European 
Southern Observatory (ESO), under the ESO-ON agreement to operate 
the 1.52m telescope; and Observat\'orio do Pico dos Dias, operated 
by the Laborat\'orio Nacional de Astrof\'\i sica (LNA).}}

\author{C\'esar A. Caretta, Marcio A. G. Maia and Christopher N. A. 
Willmer\footnote {Present address: UCO/Lick Observatory, University 
of California, Santa Cruz, 95064}.}
\affil{Depto. de Astronomia, Observat\'orio Nacional/CNPq, Rua Gal. 
Jos\'e Cristino~77, 20921-400, Rio de Janeiro~-~RJ, Brazil}

\begin{abstract}
We have carried out an independent quantitative evaluation of the
galaxy component of the ``COSMOS/UKST Southern Sky Object Catalogue''
(SSC) and the ``APM/UKST J Catalogue'' (APM).
Using CCD observations our results corroborate the accuracy of the
photometry of both catalogs, which have an overall dispersion of
about 0.2~$mag$ in the range 17~$\leq b_J \leq$~21.5.
The SSC presents externally calibrated galaxy magnitudes that
follow a linear relation, while the APM instrumental magnitudes
of galaxies, only internally calibrated by the use of stellar 
profiles, require second-order corrections. 
The completeness of both catalogs in a general field falls rapidly 
fainter than $b_J =$~20.0, being slightly better for APM.
The 90\% completeness level of the SSC is reached between 
$b_J =$~19.5 and 20.0, while for APM this happens between 
$b_J =$~20.5 and 21.0.
Both SSC and APM are found to be less complete in a galaxy cluster 
field, where completeness reachs 90\%, respectively, in the ranges 
$b_J =$~19.0-19.5 and $b_J =$~19.5-20.0.
Galaxies misclassified as stars in the SSC receive an incorrect 
magnitude because the stellar ones take saturation into
account besides using a different calibration curve.
In both cases, the misclassified galaxies show a large diversity of
colors that range from typical colors of early-types to those of
blue star-forming galaxies.
A possible explanation for this effect is that it 
results from the combination of low sampling resolutions with 
properties of the image classifier for objects with characteristic
sizes close to the instrumental resolution.
We find that the overall contamination by stars misclassified as
galaxies is $<$ 5\%  to $b_J =$~20.5, as originally estimated for 
both catalogs. 
Although our results come from small areas of the sky, they are
extracted from two different plates and are based on the comparison 
with two independent datasets. 
We conclude that both the SSC and APM can be a particularly valuable 
resource for extragalactic studies in the Southern Hemisphere once 
their limitations are taken into account.
\end{abstract}

\keywords{catalogs - galaxies: fundamental parameters (classification, 
photographic magnitudes) - techniques: image processing}

\section{Introduction}

The field of Observational Cosmology has benefited from the large 
galaxy catalogs compiled as result of high-speed machine measurements 
of photographic plates, such as the catalogs generated using the
APM (\cite{M90a}) and COSMOS (\cite{HCM89}) machines. 
However, the large areas of sky are somewhat offset by the limited 
dynamic range and inhomogeneities of the emulsions, the need for a
systematic control of the developing process, as well as by the 
limitations due to the digitization procedure. \cite{MFS95}, for 
instance, presented evidence of systematic errors in the photographic 
photometry performed by COSMOS and APM, which are important for objects 
with high surface brightness.
 
In the course of a survey we are carrying out to study the SC-16
and SC-17 superclusters identified by Abell (1961), 
we have found other systematic effects which are present in the 
COSMOS/UKST Southern Sky Object Catalogue (hereafter SSC; \cite{Y92}; 
\cite{DBE95}) and in the APM/UKST J Object Catalogue (hereafter APM; 
\cite{M90a})\footnote{available, respectively, at \\
{\it http://www.aao.gov.au/local/www/surveys/cosmos/} \ \ and 
{\it http://www.ast.cam.ac.uk/$\sim$apmcat/}}.
This effect is a systematic loss of bona-fide galaxies at
fainter magnitudes, when compared to other catalogs derived
from photographic photometry using the same plate material, and
independent CCD photometry.
This paper is divided as follows: in Section 2 we describe the data 
acquisition and catalogs; Section 3
shows the results from the comparison between the SSC and APM with
catalogs derived from PDS scans and the EIS (ESO Imaging Survey) data;
Sections 4 and 5 conclude the text, respectively with a discussion 
and a summary.

\section{The Data}

\subsection{COSMOS Catalog}

The first catalog we consider in this work is the SSC, derived from the
COSMOS scans (\cite{MS84}) of glass copies of the UKST/SERC J band 
survey plates, which is available on-line from the Anglo-Australian 
Observatory. 
The slit size used in these scans was of 16${\mu}m~\times$~16${\mu}m$, 
corresponding to 1.08\arcsec~$\times$~1.08\arcsec \ on the sky 
(see Table 1 for a summary of slit sizes and seeing for all 
catalogs). Object detection and classification were 
performed with the basic COSMOS algorithms (\cite{MS84}; \cite{TMR84};
\cite{HCM89}; \cite{BMT90}), using a threshold of 2 times the background 
standard deviation ($\sigma$), which corresponds
to a surface brightness limit of about 25.5~$b_J/\Box $\arcsec,
and considering an object whenever it contains more than 4 pixels.
The instrumental magnitudes of SSC stars are corrected for saturation 
using a function that takes into account the area of the object and its 
position on the plate.
The area is considered because it is directly related to the stellar 
magnitude ({\it e.g.} \cite{BI84}) and is independent of saturation 
(except for very bright stars, in which diffraction spikes and ghost 
halos appear).
The object position is also taken into account because the saturation 
level is subject to field effects, such as geometrical vignetting and 
intrinsic changes in sensitivity of the emulsion across the plate 
(\cite{H88}).
As described by \cite{Y92}, the SSC had two procedures for the
photometric calibration. Stars were directly calibrated from the Guide
Star Photometric Catalogue, while galaxies were
calibrated using existing CCD photometry in the B and V bands.
The uncertainty in the SSC magnitudes is estimated to be 
$\sim$~0.3-0.5~$mag$ (\cite{UHP93}), and that on positions of 
less than 0.5\arcsec \ (\cite{DBE95}).
Previous estimates for SSC completeness were of about 95\% at
$b_J =$~20.0 ({\it e.g.} \cite{HCM89}, \cite{L97}) or $b_J =$~20.5 
({\it e.g.} \cite{N92}, \cite{EM95}).

\subsection{APM Catalog}

The second catalog used in this analysis is the APM, produced from
APM machine scans (\cite{K84}) of similar glass copies of the 
UKST/SERC J band survey, which is partially available on-line from
the Institute of Astronomy - Cambridge, UK. 
The APM also used a pixel size of 16${\mu}m~\times$~16${\mu}m$
although with a sampling interval (steps) of half this size,  
resulting in a 0.54\arcsec~$\times$~0.54\arcsec \ sampling rate.
Objects were detected with the APM image analysis algorithm
(\cite{K84}) and classified using the profile technique, which is
based on the combination of up to 8 isophotal areas, the peak 
intensity and the radius of gyration (\cite{M88a}, \cite{M90a}).
The adopted detection threshold is 2$\sigma$ above the local sky 
background intensity, which corresponds to an isophotal limit of 
about 25 $b_J/\Box $\arcsec, with a  minimum area for each object 
of 16 pixels (\cite{M88b}).
The isophotal magnitudes are corrected for 
position-dependent field effects for each plate 
(geometrical vignetting and differential desensitization).
These corrected instrumental magnitudes are then internally 
calibrated by the use of the stellar profiles (M. Irwin, private 
communication). 
Corrections on a plate to plate basis are calculated
and added to the instrumental magnitudes so that the final 
ones are uniform throughout the survey (\cite{M90b}). 
The internal accuracy of APM magnitudes was estimated to be about 
0.1-0.2~$mag$ up to $b_J \sim$~20.5 (\cite{M90a}), while externally 
they were estimated to be accurate to 0.3~$mag$, becoming worse 
towards the brighter side (\cite{I99}). This is due to the use of
the calibration based on stellar profiles, which tends to
make galaxy magnitudes artificially brighter.
The APM positions have internal uncertainties of 0.1\arcsec \ and 
external ones of 0.5\arcsec \ (\cite{I99}).
Completeness estimates were of 90-95\% and contamination of 5-10\%
up to $b_J =$~20.5 (\cite{M90a}).

\subsection{PDS Catalog}

An ESO on-film copy of the UKST/SERC IIIa-J plate (field 535) was 
digitized with the PDS 1010A microdensitometer of the Observat\'orio 
Nacional using a square pixel size and  step of 10${\mu}m$, which 
corresponds to a sampling rate of 0.67\arcsec. The FWHM of stellar
objects measured on the film is of about 2.3\arcsec.
The scanned region comprised an area of 30\arcmin~$\times$~30\arcmin \ 
(0.25$\Box $\arcdeg) centered at $\alpha_{2000} = 23^h 07^m 33^s$ and 
$\delta_{2000} =$ -22\arcdeg 33\arcmin, containing two neighboring
galaxy clusters, A2534 and A2536.     
The FOCAS package (\cite{JT81}, \cite{V82}) was adopted to perform the 
object detection and star/galaxy classification.  
The threshold level was of 3$\sigma$ above the local sky background 
intensity, corresponding to an isophotal limit of about 
25.0~$b_J/\Box $\arcsec. 
An object was considered only if it comprised a minimum of 12 pixels. 
The absolute calibration was done using B and V CCD galaxy magnitudes 
from images obtained at the 1.60m telescope of the Observat\'orio do 
Pico dos Dias (OPD), Bras\'opolis, Brazil. 
A sequence of 26 galaxies located in the central region of A2534 had 
aperture magnitudes measured, using IRAF\footnote{IRAF is distributed 
by the National Optical Astronomy Observatories (NOAO) which is operated 
by the Association of Universities for Research in Astronomy, Inc. under
contract to the National Science Foundation;}/DAOPHOT package, spanning 
from $b_J =$~18.0~to~21.5. 
We have chosen a 4\arcsec \ radius aperture as a compromise in order 
to avoid a significant loss of light for the brighter objects,  while
minimizing the background noise for the fainter ones. 
Similar aperture magnitudes were measured in the FOCAS catalog. 
The final calibration of the scan data (both stars and galaxies) 
was performed using the color equation obtained  by \cite{BG82}:

\begin{equation}
b_J = B - 0.28 ( B - V )
\end{equation}

\noindent where our B and V CCD magnitudes are in the standard 
Johnson-Cousins photoelectric system, the passbands used in our CCD 
observations being in agreement with those tabulated by \cite{B90}. 
The uncertainty in our calibrated $b_J$ magnitudes is estimated to be 
about 0.15~$mag$. The total magnitudes that were finally used from this 
catalog were measured with FOCAS growing area process, which
considers the surface brightness inside an object area 20\% larger
(\cite{JT81}).
We should note that our stellar magnitudes are not strictly correct
brighter than $b_J =$~19.0 since no saturation correction has been 
applied and the calibration used for all objects was that determined 
for galaxies.

In order to estimate the reliability of the detection and classification 
of images by FOCAS, we compared part of our PDS catalog with a CCD
derived one that
will be described below as the Supplementary Catalog 1.
This comparison allowed us to estimate that the FOCAS classification is 
reliable (completeness~$\sim$~95\%) for galaxies with magnitudes up to 
$b_J =$~21.5.

\subsection{EIS Catalog}

A fourth catalog, also taken as an external and independent check,
is that derived from multicolor CCD imaging for the ESO Imaging
Survey (EIS, \cite{N98}). 
In particular, we used the catalog for Patch B which is located at 
the Southern Galactic Pole ($\alpha_{2000} = 00^h 51^m 00^s$ and  
$\delta_{2000} =$ -28\arcdeg 54\arcmin 00\arcsec, field 411), 
in B and V bands (\cite {P98}).
The EIS frames have a pixel size of 0.27\arcsec, and the median 
seeings are 1.2\arcsec \ and 0.9\arcsec, respectively for B and V 
bands.
The Patch B is composed by a sequence of 150 overlapping images of 
9\arcmin $\times$ 8.5\arcmin, from which we selected 32 
unintersecting frames. 
The corresponding area is of 0.68$\Box $\arcdeg.
Object detection and classification were performed with SExtractor 
(\cite {BA96}), using a 0.6$\sigma$ threshold (corresponding to 
$\sim$~26.8~$B/\Box $\arcsec \ and 
$\sim$~26.4~$V/\Box $\arcsec \ surface brightness limits) and a 
back-propagation neural-network, fed with isophotal areas and
peak intensity of the profiles, for star/galaxy separation.
EIS data were calibrated directly from Landolt (1992 a,b) standard 
stars, for frames observed in photometric conditions. Both Patch B 
catalogs are about 95\% complete in detection to B~=~V~=~23.0 
(\cite {P98}).
We adopted a stellarity index of 0.75 to separate stars from 
galaxies in the EIS catalogs, and the final completeness limit
is estimated to be B~=~V~=~22.0.
The transformation from B and V EIS magnitudes to the $b_J$
system was done using Equation 1. The original B and V EIS 
magnitudes were first converted to Johnson-Cousins system using the 
relation presented in Prandoni {\it et al.} (1998):

\begin{equation}
B = B_{EIS} + 0.161 ( B - V )_{EIS}
\end{equation}
\begin{equation}
V = V_{EIS} - 0.057 ( B - V )_{EIS}
\end{equation}

Down to  $b_J =$~21.5, about 90\% of the B-band objects had a
corresponding match in the V-band data, while the classification was 
coincident in more than 95\% of the cases.

\subsection{Supplementary Catalogs}

Two other catalogs were used as supplementary ones, both comprising
only the PDS area. 

The first Supplementary Catalog was generated from a set of 
3 CCD images obtained with the OPD 1.60m telescope, exposed with no 
filters (aiming to get more light in less time) under a
$\sim$~1\arcsec\ seeing.
These images targeted the central parts of the clusters A2534 and
A2536. Each image covers an area of
3.9\arcmin~$\times$~5.6\arcmin, while 
the pixel size of these frames corresponds to 0.29\arcsec.
This catalog was obtained using FOCAS, and was only used in this work
as an additional and better quality source for the classification of
objects in the area measured with the PDS.

The second Supplementary Catalog is in the R-band, derived from
an ESO Schmidt Telescope plate, which was digitized and processed with
the APM. This catalog was calibrated using 14 galaxies that have CCD R
magnitudes, 5 measured at the OPD 1.60m telescope and 9 from
\cite{CW94}. This catalog was used to obtain the
$b_J - R$ color index for objects in the PDS region.

\section{Comparison between catalogs}

\subsection{Number counts of galaxies on PDS area (field 535)}

The number counts of galaxies on PDS area, in the three $b_J$-band 
catalogs (PDS, SSC and APM), are shown on Table 2 and panel $(a)$ 
of Figure 1. Table 2 also shows the number counts for stars and
the total numbers of detected objects, including objects classified
as ``faint'' (on SSC) and ``merged'' (on APM).
Because SSC objects fainter than $b_J \sim $~21.5 are no longer 
separated into galaxies and stars, being simply classified as 
``faint'', the number counts of galaxies and stars are small in the 
last magnitude bin in the table (21.5~$\le b_J <$~22.0).

Considering the total counts to $b_J =$~21.5, the PDS and SSC show
about the same number of objects, indicating that they achieve a 
similar efficiency in detection. The APM, otherwise, has lower counts,
but this is a consequence of a zero-point difference, which will be
described in more detail in Section 3.3. By correcting the 
magnitudes of APM catalog for this offset, this means taking the 
counts up to $b_J =$~22.0, all the three catalogs show roughly 
the same total counts. 
However, the behavior of the galaxy and star counts are quite 
different in each catalog. For the PDS, the total number of galaxies 
is more than twice that of the SSC and APM. Also the galaxy differential 
counts increase about exponentially for this catalog till the
limit considered, as expected from its estimated 
completeness (see Figure 1).
For the SSC, on the other hand, the differential galaxy  
counts reach a maximum in the bin 20.0~$\le b_J <$~20.5, 
which may indicate that the successful classification of 
objects does not reach much beyond this limit.
The most intriguing counts are those of APM: they are flatter than
the others, having a larger number of galaxies in the brighter 
bins and a smaller number in the fainter ones, as compared to the 
other two catalogs. This effect may be explained, for the brighter 
side, by the use of the magnitude calibration for stars as mentioned 
in Section 2.2, where galaxy magnitudes are made artificially brighter.
Another possible contribution, more effective for the fainter side, 
may be due to the fact that the APM
catalog that we use for this region was not derived from scans
of field 535, but from the adjacent field 604, since the 
original plate was not available at the time this work was
being carried out.
In this northern plate the PDS region is located close to its 
edge, and some photometric and/or classification effects might still 
be present (\cite {I99}), despite the corrections applied to APM 
objects as described in section 2.2.

\subsection{Number counts of galaxies on EIS area (field 411)}

To ensure that the differing number counts found in the previous 
section are not an effect unique to the PDS Catalog, we made an 
independent check 
on the SSC and APM, using the EIS Catalog, which probes a
different part of the sky, reachs a fainter magnitude limit 
and has overall a better classification than the other 
catalogs, due to the quality and sampling rate of the CCD data.
The number counts for the EIS field are shown on Table 3 and 
panel $(b)$ of Figure 1, 
which shows, at the faint end, that they are well reproduced by a  
straight line of slope 0.49 in logarithmic scale, as expected for 
field galaxies in this magnitude range ({\it e.g.} \cite{MSFJ91}). 
Again, SSC counts have a maximum in the bin 20.0~$\le b_J <$~20.5
while APM has flatter counts in the bright end. From the figure, 
one can see that APM also presents lower counts for $b_J$ fainter 
than 20.0. This effect is unlikely to be caused by the 
superestimation of magnitudes described above, as the stellar 
saturation correction becomes very small at these magnitudes.

\subsection{Distribution of magnitudes}

The comparison between magnitudes derived from the SSC and APM catalogs 
with those of the EIS are displayed respectively in panels $(a)$ and 
$(b)$ of Figure 2.
In these panels objects classified in both catalogs as galaxies are 
represented by solid circles, those classified as stars, by points, and 
those for which the classifications disagree, by crosses. The figure
shows that each object type occupies a different {\it locus}
according to its match.

Polynomial fits to the distributions in Figure 2$a$ are displayed in 
Figure 2$c$, as well as the median values and rms of the distributions 
for each magnitude bin.
The overall rms are 0.19, 0.22 and 0.28, respectively for galaxies in 
both (solid line), stars in both (dotted line) and EIS galaxies 
classified as stars in SSC (dashed line). Due to the correction 
applied to stellar magnitudes in SSC, described on section 2.1, the 
distribution of magnitudes for stars was expected to be linear, 
meanwhile it is best represented by a second order polynomial fit, 
implying that the saturation correction is not entirely effective. 
The distribution of galaxy magnitudes, on the other hand, is very 
well represented by a linear relation.
The relation between the SSC and CCD galaxy magnitudes, in the range 
17.0~$\le b_J \le$~21.5, is:

\begin{equation}
b_J^{(EIS)} = -0.43 + 1.027 \ b_J^{(SSC)}
\end{equation}

The different inclinations and zero-points between both distributions
may be explained by the presence/absence of the saturation correction 
and the distinct calibration procedures.
The behavior of the magnitude distribution of EIS galaxies
classified as stars in SSC is completely distinct from the other two. 
This effect may be explained by the fact that galaxies have larger
areas than a star of the same magnitude, yet receive the
area-dependent magnitude correction as stars, so that the 
correction they are subjected to is excessively large in spite of the 
fact that they practically do not have saturated pixels.

The panel $(b)$ of Figure 2 shows the comparison between 
EIS and APM. As for the SSC, the APM magnitudes also received a
correction to linearize stellar magnitudes based on their profiles.
However, the correction was applied to the instrumental 
magnitudes of all objects, including galaxies.
A direct consequence of this is that the galaxies do not follow
a linear relation, with their {\it locus} bending towards 
the brighter APM magnitudes. 
The EIS galaxies classified as stars in APM follow the same distribution 
above but, as they appear at fainter magnitudes, it is not as evident 
as with the SSC.
We also show in panel $(d)$ of Figure 2 the polynomial fits to these 3
distributions.
The overall dispersions are 0.24, 0.13 and 0.22, 
respectively for galaxies in both, stars in both and EIS galaxies 
classified as stars in APM.  The relation of APM {\it versus} EIS 
galaxy magnitudes, obtained in the range 17.0~$\le b_J \le$~21.5, 
is given by the expression:

\begin{equation}
b_J^{(EIS)} = 27.21 - 1.476 \ b_J^{(APM)} + 0.0566 \ [b_J^{(APM)}]^2
\end{equation}

When compared to PDS, the SSC and APM magnitudes show similar behavior,
except that the stars in PDS are not corrected for saturation and so
their distribution is curved in the same way as the one for the galaxies
in panel $(b)$ of Figure 2, only with a shallower inclination. Another 
effect found in this last comparison is that the APM magnitudes for the 
PDS area (extracted from the edge of field 604 instead of field 535)
have a shift, that is of about 0.5~$mag$ around $b_J =$~21.5. 
So, returning to Table 2, the number counts that should 
be considered for the comparison with PDS and SSC is the one at 
$b_J =$~22.0.

\subsection{Completeness estimates}

Since we have mapped the magnitude differences between catalogs and 
defined the zero points, we are able to estimate the real completeness 
and contaminations of SSC and APM, in the areas we studied, based on 
the comparison of these catalogs with EIS and PDS.
These estimates are shown in Table 4 and Figure 3. 
The $\sim$~5\% of EIS objects with non-coincident B and V 
classification were excluded to guarantee the reliability in the 
EIS catalog classification.
In addition, we should note that slightly less than 5\% of the objects 
in the other two catalogs may have been lost in the matching procedure. 
From the table and figure we note that both SSC 
and APM start to loose galaxies in the brighter bins, although 
completeness estimates for $b_J <$~19.0 do not have enough 
statistical weight because of the small number of objects.
Fainter than $b_J =$~19.5, the SSC completeness drops quickly, reaching 
a level of 90\% in the bin 19.5~$\le b_J <$~20.0, and 65\% around 
$b_J =$~21.5. The APM completeness is less steep, reaching the 90\% 
level at least 0.5~$mag$ deeper than SSC.

Although the area used in the comparison with EIS is almost twice as 
large as that used with the PDS, the latter contains two clusters 
at $z \approx 0.2$, therefore a higher galaxy density (see also the 
bump on galaxy counts in Figure 1$a$). Because of this the 
completenesses of the SSC and APM relative to PDS are smaller than 
when considered relative to the EIS. 
The 90\% completeness level is reached by the SSC in the range 
$b_J =$~19.0-19.5, and $b_J =$~19.5-20.0 by the APM. 
As clusters tend to concentrate more early-type galaxies than the 
general field and because this type of galaxy may be more easily 
confused with stars, this implies that the efficiency of image 
classification by SSC and APM could be less successful in such fields.

Our results show that as far as the contamination by misclassified 
objects is concerned, both catalogs have small fractions 
of stars classified as galaxies, $<$~3\% up to the adopted magnitude
limit. However, as the regions we examined are located at a high
galactic latitudes, where the star density is small, it is possible
that this contamination rate may increase as one goes towards lower
galactic latitudes.

\subsection{What are the misclassified SSC and APM objects?}

We present in Table 5 and Figure 4 a subsample of the brightest
objects with conflicting classifications that are contained in
both the PDS and SSC catalogs.
The images displayed in Figure 4 are from the PDS digitization of 
the on-film copy of the UKST plate and contained in a small SSC 
magnitude range ($18.4 < b_J < 18.7$).
The objects in the table and figure are organized according to their 
classification in the catalogs as follows: those classified as galaxies 
in both (objects 1-5); those classified as galaxies in PDS and as 
stars in SSC (6-10); and finally, those which are stars in both (11-15). 
Concerning the classification, objects 1-5 (first column of Figure 4) are 
obvious galaxies, while 11-15 (third column of figure) actually look like 
stars.  The objects 6-10 (in the middle column of figure) cannot be 
readily classified by eye, although they show a flatter surface 
brightness profile, more typical of galaxies.
Some of the objects in this table were observed spectroscopically, and
measured redshifts are listed in column 8 of Table 5. Details about 
these observations and the remaining data will be presented elsewhere. 
For 2 of the SSC ``stars'', which were classified as ``galaxies'' in 
the PDS Catalog, the spectra reveal that the PDS classification is 
correct. The spectra for these galaxies are displayed in Figure 5.

In order to characterize further properties of the galaxies 
misclassified as stars in SSC and APM, we separated objects with 
conflicting classifications into three groups: those classified as 
stars only in SSC; those classified as stars only in APM; and those 
classified as stars in both. In principle, this last group may have 
some contamination from real stars misclassified as galaxies by PDS. 
The results can be seen in Figure 6, where $b_J$ magnitudes are plotted
against FOCAS core magnitudes, measured from the brightness of the 9 
central pixels: PDS against SSC in panel ($a$) and PDS against APM in 
panel ($b$). The PDS galaxies classified as stars in both
SSC and APM (plotted with a open square symbol) occupy a region slightly
above the stellar {\it locus} (represented by dots) in both panels. 
On the other hand, PDS galaxies classified as stars in only one of the
catalogs (crosses) occupy the {\it locus} of real galaxies 
(solid circles).   

The ($b_J-R$) color distribution for the five sub-groups (galaxies in 
all catalogs, stars in all, PDS galaxies classified as stars in both 
APM and SSC, and PDS galaxies classified as stars in either APM or SSC) 
is presented in Figure 7.  
The magnitude interval is restricted to 19 $\le b_J \le$ 21 to 
minimize effects of saturation and detection incompleteness. 
The probability of each sub-group pair being derived from the same
parent distribution ($P_{KS}$) was calculated through the
Kolmogorov-Smirnov (KS) test, and  the results are summarized
in Table 6, which identifies the samples, the number of members in
each sample, the KS coefficient and $P_{KS}$ values. 
PDS galaxies classified as stars only by the SSC or the APM catalogs,
have a good chance of being galaxies.  
The PDS galaxies classified as stars by both SSC and APM have a 21\% 
probability level of being actual stars.

For some of the PDS galaxies that were misclassified as stars by
the SSC and/or APM, we have CCD images. A representative subsample of
these is presented in Figure 8, while the parameters are presented 
in Table 7.
The left hand column of Figure 8 shows the PDS images while the right
hand side shows the white light CCD images for the same object. 
Also shown are the FOCAS isophotes which due to a plotting artifact
are shifted 1 pixel upwards relative to the object images.
The visual inspection shows that up to $b_J =$~21, all PDS  galaxies
classified  as stars in either SSC or APM seem to be real galaxies.
A similar analysis for PDS galaxies classified as stars in both SSC and 
APM, shows that about 70\% are indeed real galaxies in CCD frames, 
while the remaining 30\% resemble stars.

By obtaining spectra for all objects in the range 
16.5~$< b_J <$~20.2, in an area containing the Fornax cluster field, 
\cite{D99} identified some compact galaxies that were 
classified as stars in the APM catalog. \cite{D99} also showed
that these galaxies tend to have blue colors, high luminosities and
strong emission lines.
In the case of galaxies misclassified as stars in the SSC,
although part may be indeed this kind of object, compact
galaxies cannot account for all. The two objects 
for which we have spectra do indeed present emission lines (Figure 6),
but which cannot be considered as being particularly strong. The 
luminosities of these galaxies are in general smaller than the 
\cite{D99} compacts.
The galaxies misclassified in the SSC have colors that 
are spread over the range  (0.5~$\lesssim b_J-R \lesssim$~2.5), 
while the \cite{D99} galaxies are, in general, bluer 
(0.2~$\lesssim b_J-R \lesssim$~1.4).

\section{Discussion}

From the analyses presented in the previous section, we conclude that 
there exists a smaller success rate in the classification for faint 
objects by the high speed machines COSMOS and APM relative to catalogs
derived from CCD photometry and from PDS  scans.
More specifically both the SSC and the APM begin to loose galaxies 
around $b_J =$~19.5.

As the PDS data are extracted from the same material - the fact that 
SSC and APM come from glass copies of the original plates and PDS 
Catalog comes from a film one does not affect significantly the 
instrumental resolution - we are lead to suppose that the discrepancy 
in efficiency may be related to the digitization process and/or in 
the classification algorithms/procedures.
Concerning the digitization, the different resolutions may partially 
explain the lower efficiency of SSC, since the effective resolution
(pixel size combined with sampling rate) used for this catalog is 
1.5 times lower than that of the PDS and 2 times lower than the APM.
Another effect that could explain the misclassified galaxies  may be 
related to the classification procedures.  The SSC, for example, uses 
the so called ``parametric'' classifier, which takes into account the 
scatter of certain combination of parameters, such as the area and the 
maximum intensity, with the magnitude.
On the other hand, the classifications of the PDS Catalog come from 
FOCAS, which uses the ``resolution'' classifier of Valdes (1982). 
This classifier is based on the point spread function (PSF) of stellar 
profiles in the image, and in the case of the PDS scans, the PSF was 
optimized specifically for this region of the plate.
Our results are thus in accordance with those of \cite{WP92} who found 
that the ``resolution'' approach achieves a better performance than the 
``parametric'' used by the SSC (and COSMOS).  The profile classification 
of APM may also be considered as a ``parametric'' procedure.

Even though the digitization resolution seems to be the dominant
factor to jeopardize the classification efficiency, we are 
not able to produce a controlled experiment, involving all the 
processes on the same data, in order to determine the relevance of 
the instrumental resolution and the classification algorithms.

\section{Summary}

We have compared high speed machine based catalogs which are available
on-line, SSC and APM, to catalogs generated from a slow machine, PDS, 
and from CCD data, EIS. Each of the catalogs used independent 
algorithms for detection and classification of object images.

The comparison with the PDS catalog gave the qualitative results: 
the scanning resolution and/or the nature of the classification 
algorithm may determine the efficiency of the machine to separate 
galaxies and stars. With the same photographic material, PDS 
0.7\arcsec \ sampling rate plus FOCAS ``resolution'' classifier 
achieved a better performance than SSC 1.1\arcsec \ sampling rate 
plus ``parametric'' classifier.  APM, although not tested for the 
same plate material, with a 0.5\arcsec \ sampling rate plus a 
classifier similar to the ``parametric'' one, achieved intermediate 
results.  

The comparison with EIS is the most robust and gives us some 
quantitative results. Concerning the galaxy photometry, we find 
that calibrated data from the SSC have a linear magnitude system 
and a small dispersion of about 0.2~$mag$.  The APM data also 
present this small dispersion but, as they are not calibrated 
photometrically and have a stellar-source correction applied, 
they cannot be taken at face value. We find that APM galaxy 
magnitudes do not follow a linear relation and require second 
order corrections. 
The SSC catalog  is 90\% complete for galaxies down to 
$b_j =$~19.5-20.0, beyond which the completeness drops rapidly. 
The APM catalog is 90\% complete for galaxies down to 
$b_j =$~20.5-21.0.
These completeness become slightly lower when a cluster field is 
considered due to the higher proportion of early-type galaxies, 
which, because of their steeper profiles, are more likely to be 
misclassified as stars. The misclassified galaxies encompass a 
large color range (0.5~$\leq B_J-R \leq$~2.5), which includes 
typical colors of blue high-luminosity star-forming compacts to 
those of normal early-type galaxies. 
Our results also show that for the high galactic latitude fields 
we considered, the proportion of stars misclassified as galaxies 
is less than 3\% up to the adopted magnitude limit.  

Although these results were obtained in two small areas of the sky, 
they demonstrate that both the SSC and APM catalogs are valuable 
resources for extragalactic research once their limitations are 
taken into account.

\acknowledgments

We would like to thank the staff and night assistants of OPD/LNA, 
CASLEO and ESO; the AAO for providing the SSC and the IoA/Cambridge 
for providing the APM. 
We are also grateful to G. Pizzaro for taking our ESO plates, 
M. Irwin for providing the digitization; P.S.S. Pellegrini for 
valuable suggestions and discussion and H.T. Mac Gillivray for helpful 
comments. We also acknowledge the anonymous referee whose comments and 
suggestions improved a previous version of this paper. 
This research has also made 
use of NASA/IPAC Extragalactic Database (NED). The authors acknowledge 
use of the CCD and data acquisition system supported under U.S. 
National Science Foundation grant AST 90-15827 to R.M. Rich. C.A.C. 
acknowledges financial support from CNPq and CAPES scholarships, 
M.A.G.M. to CNPq grant 301366/86-1, and C.N.A.W. to CNPq grant 
301364/86-9 and NSF AST 95-29028.

\newpage

\pagestyle{empty}
\makeatletter
\def\jnl@aj{AJ}
\ifx\revtex@jnl\jnl@aj\let\tablebreak=\nl\fi
\makeatother

\begin{deluxetable}{llcllcr}
\small
\tablecolumns{7}
\tablenum{1}
\tablewidth{0pt}
\tablecaption{Catalogs used in this work}
\tablehead{
Catalog & Source & Band & Digitization & Reduction\tablenotemark{\ast} & 
Sampling Rate & Seeing}
\startdata 
SSC & UKST/ESO plate & $b_J$    & COSMOS  & COSMOS      & 1.08" & $<$ 3"\tablenotemark{\ddagger} \nl
APM & UKST/ESO plate & $b_J$    & APM     & APM         & 0.54" & $<$ 3"\tablenotemark{\ddagger} \nl
PDS & UKST/ESO film  & $b_J$    & PDS     & FOCAS       & 0.67" & 2.3"\tablenotemark{\dagger} \nl
EIS & NTT CCD        & B,V      & \nodata & SExtractor  & 0.27" & 1.2"/0.9" \nl
Suppl1 & OPD CCD     & white light & \nodata & FOCAS    & 0.29" & 1.0"  \nl
Suppl2 & ESO Schmidt plate & R  & APM     & APM         & 0.54" & 1.5" \nl
\enddata
\tablenotetext{\ast}{ Object detection and classification.}
\tablenotetext{\ddagger}{ Tipical seeing for UKST survey plates 
(Heydon-Dumbleton, Collins \& MacGillivray 1989).}
\tablenotetext{\dagger}{ FWHM of stellar profiles inside PDS area.}
\end{deluxetable}

\pagestyle{empty}
\makeatletter
\def\jnl@aj{AJ}
\ifx\revtex@jnl\jnl@aj\let\tablebreak=\nl\fi
\makeatother

\begin{deluxetable}{ccrrrcrrrcrrr}
\small
\tablecolumns{13}
\tablenum{2}
\tablewidth{0pt}
\tablecaption{Number counts for PDS area (field 535 - 0.25$\Box $\arcdeg)}
\tablehead{Range of $b_J$ & \colhead{} & \multicolumn{3}{c} {PDS} & 
 \colhead{} & \multicolumn{3}{c} {SSC} & \colhead{} & 
 \multicolumn{3}{c} {APM\tablenotemark{\dagger}} \nl
\cline{3-5} \cline{7-9} \cline{11-13}
\tablevspace{0.05cm}
magnitudes\tablenotemark{\ast} & & gals. & stars &
 total\tablenotemark{\ddagger} & & gals. & stars &
 total\tablenotemark{\ddagger, \$} & & gals. & stars &
 total\tablenotemark{\ddagger, \sharp} }
\startdata 
$\rightarrow$ 17.0 & &   0 &  23 &   23 &  &   0 & 128 &  128 &  &   9 &  81 &   90 \nl
17.0 - 17.5        & &   1 &  34 &   58 &  &   1 &  27 &  156 &  &   7 &  19 &  128 \nl
17.5 - 18.0        & &   4 &  40 &  102 &  &   1 &  39 &  196 &  &   3 &  24 &  158 \nl
18.0 - 18.5        & &  11 &  50 &  163 &  &   6 &  37 &  239 &  &  13 &  20 &  195 \nl
18.5 - 19.0        & &  20 &  55 &  238 &  &  15 &  40 &  294 &  &  13 &  32 &  242 \nl
19.0 - 19.5        & &  34 &  41 &  313 &  &  35 &  69 &  398 &  &  15 &  35 &  300 \nl
19.5 - 20.0        & &  93 &  47 &  453 &  &  70 &  84 &  552 &  &  55 &  36 &  399 \nl
20.0 - 20.5        & & 122 &  59 &  634 &  &  93 & 115 &  760 &  &  58 &  79 &  546 \nl
20.5 - 21.0        & & 143 &  67 &  844 &  &  64 & 135 &  959 &  &  72 &  77 &  700 \nl
21.0 - 21.5        & & 260 &  99 & 1203 &  &  41 & 177 & 1177 &  &  58 & 139 &  900 \nl
21.5 - 22.0        & & 308 & 146 & 1657 &  &   8 &   0 & 1356 &  &  65 & 168 & 1137 \nl
\tablevspace{0.05cm}
\cline{1-13}
\tablevspace{0.05cm}
TOTALS             & & 996 & 661 & 1657 &  & 334 & 851 & 1356 &  & 368 & 710 & 1137 \nl
\enddata
\tablenotetext{\ast}{ Each catalog is considered in its original magnitude system.}
\tablenotetext{\dagger}{ Extracted from field 604.}
\tablenotetext{\ddagger}{ The totals are cumulative.}
\tablenotetext{\$}{ Including objects classified as ``faint''.}
\tablenotetext{\sharp}{ Including objects classified as ``merged''.}
\end{deluxetable}

\pagestyle{empty}
\makeatletter
\def\jnl@aj{AJ}
\ifx\revtex@jnl\jnl@aj\let\tablebreak=\nl\fi
\makeatother

\begin{deluxetable}{ccrrrcrrrcrrr}
\small
\tablecolumns{13}
\tablenum{3}
\tablewidth{0pt}
\tablecaption{Number counts for EIS area (field 411 - 0.68$\Box $\arcdeg)}
\tablehead{Range of $b_J$ & \colhead{} & \multicolumn{3}{c} {EIS} & 
 \colhead{} & \multicolumn{3}{c} {SSC} & \colhead{} & 
 \multicolumn{3}{c} {APM} \nl
\cline{3-5} \cline{7-9} \cline{11-13}
\tablevspace{0.05cm}
magnitudes\tablenotemark{\ast} & & gals. & stars &
 total\tablenotemark{\ddagger, \dagger} & & gals. & stars &
 total\tablenotemark{\ddagger, \$} & & gals. & stars &
 total\tablenotemark{\ddagger, \sharp} }
\startdata 
$\rightarrow$ 17.0 & &    2 &  202 &  205 &  &   2 &  252 &  254 &  &   25 &  152 &  200 \nl
17.0 - 17.5        & &    0 &   50 &  256 &  &   1 &   55 &  310 &  &   10 &   47 &  262 \nl
17.5 - 18.0        & &    6 &   49 &  311 &  &   4 &   69 &  383 &  &   19 &   52 &  339 \nl
18.0 - 18.5        & &   28 &   57 &  397 &  &  25 &   74 &  482 &  &   26 &   47 &  420 \nl
18.5 - 19.0        & &   36 &   69 &  505 &  &  39 &   85 &  606 &  &   26 &   72 &  527 \nl
19.0 - 19.5        & &   53 &   81 &  643 &  &  61 &  111 &  778 &  &   53 &   98 &  690 \nl
19.5 - 20.0        & &   84 &  112 &  842 &  &  90 &  142 & 1010 &  &   84 &  130 &  922 \nl
20.0 - 20.5        & &  133 &  128 & 1113 &  & 139 &  165 & 1314 &  &  123 &  176 & 1243 \nl
20.5 - 21.0        & &  218 &  146 & 1499 &  & 132 &  192 & 1638 &  &  180 &  204 & 1631 \nl
21.0 - 21.5        & &  419 &  157 & 2128 &  & 115 &  240 & 1993 &  &  242 &  313 & 2198 \nl
21.5 - 22.0        & &  688 &  135 & 3069 &  &  14 &   37 & 2535 &  &  276 &  515 & 3006 \nl
\tablevspace{0.05cm}
\cline{1-13}
\tablevspace{0.05cm}
TOTALS             & & 1667 & 1186 & 3069 &  & 622 & 1422 & 2535 &  & 1064 & 1806 & 3006 \nl
\enddata
\tablenotetext{\ast}{ Each catalog is considered in its original magnitude 
system.}
\tablenotetext{\ddagger}{ The totals are cumulative.}
\tablenotetext{\dagger}{ Including objects with non-coincident classification on the two EIS bands.}
\tablenotetext{\$}{ Including objects classified as ``faint''.}
\tablenotetext{\sharp}{ Including objects classified as ``merged''.}
\end{deluxetable}

\pagestyle{empty}
\makeatletter
\def\jnl@aj{AJ}
\ifx\revtex@jnl\jnl@aj\let\tablebreak=\nl\fi
\makeatother

\begin{deluxetable}{ccrrcrrcrrcrr}
\small
\tablecolumns{13}
\tablenum{4}
\tablewidth{0pt}
\tablecaption{Completeness of the SSC and APM according to the PDS and EIS Catalogs}
\tablehead{Range of & \colhead{} & \multicolumn{5}{c} {SSC} 
& \colhead{} & \multicolumn{5}{c} {APM} \nl
\tablevspace{-0.05cm}
 \cline{3-7} \cline{9-13}
\tablevspace{0.05cm}
$b_J$ & \colhead{} & \multicolumn{2}{c} {$\times$PDS} & \colhead{} & 
\multicolumn{2}{c} {$\times$EIS} & \colhead{} & \multicolumn{2}{c} 
{$\times$PDS} & \colhead{} & \multicolumn{2}{c} {$\times$EIS} \nl
\tablevspace{-0.05cm}
 \cline{3-4}  \cline{6-7} \cline{9-10} \cline{12-13}
\tablevspace{0.05cm}
magnitudes & \colhead{} & diff. & integr. & \colhead{} & diff. & integr. &
\colhead{} & diff. & integr. & \colhead{} & diff. & integr.}
\startdata 
$\rightarrow$ 17.0 & & 100.0 & 100.0 & & 100.0 & 100.0 & & 100.0 & 100.0 & &  50.0 &  50.0 \nl
17.0 - 17.5        & &   0.0 &  50.0 & & 100.0 & 100.0 & & 100.0 & 100.0 & & 100.0 &  50.0 \nl
17.5 - 18.0        & &  88.9 &  80.0 & &  66.7 &  75.0 & & 100.0 & 100.0 & & 100.0 &  75.0 \nl
18.0 - 18.5        & & 100.0 &  88.9 & &  92.6 &  88.6 & & 100.0 & 100.0 & &  95.2 &  92.0 \nl
18.5 - 19.0        & & 100.0 &  94.9 & &  88.6 &  88.6 & & 100.0 & 100.0 & & 100.0 &  96.3 \nl
19.0 - 19.5        & &  87.5 &  91.1 & & 100.0 &  93.5 & &  90.3 &  95.2 & & 100.0 &  98.0 \nl
19.5 - 20.0        & &  80.0 &  84.8 & &  90.4 &  92.2 & &  84.6 &  89.0 & &  97.2 &  97.6 \nl
20.0 - 20.5        & &  61.8 &  75.6 & &  75.9 &  85.8 & &  72.3 &  82.0 & &  93.0 &  95.8 \nl
20.5 - 21.0        & &  34.4 &  61.8 & &  61.6 &  76.5 & &  51.0 &  70.7 & &  83.9 &  90.9 \nl
21.0 - 21.5        & &  11.8 &  48.3 & &  40.5 &  64.0 & &  31.5 &  56.4 & &  69.5 &  81.4 \nl
\enddata
\end{deluxetable}

\pagestyle{empty}
\makeatletter
\def\jnl@aj{AJ}
\ifx\revtex@jnl\jnl@aj\let\tablebreak=\nl\fi
\makeatother

\begin{deluxetable}{cccccccccccccc}
\small
\tablenum{5}
\tablecolumns{14}
\tablewidth{0pt}
\tablecaption{Examples of objects in PDS area}
\tablehead{\colhead{} & \colhead{} & \multicolumn{2}{c} {Coordinates} 
& \colhead{} & \multicolumn{2}{c} {Classification\tablenotemark{\sharp}} 
& \colhead{} & \multicolumn{2}{c} {b$_J$ magnitude} & \colhead{} 
& \colhead{} & \colhead{} & \colhead{} \nl
 \cline{3-4} \cline{6-7} \cline{9-10}
\tablevspace{0.05cm}
Nr. & \colhead{} & $\alpha_{2000}$ & $\delta_{2000}$ & \colhead{} & PDS 
& SSC & \colhead{} & PDS & SSC & \colhead{}  
& $a \times b$ \tablenotemark{\ast} & $z$ & Obs.\tablenotemark{\dagger}}
\startdata 
 1 & & 23 06 41.25 & -22 53 10.2 & & g & 1 & & 18.06 & 18.48  & & 8.3 $\times$ 3.7 &  - & - \nl
 2 & & 23 07 44.59 & -22 31 08.9 & & g & 1 & & 18.09 & 18.52  & & 7.3 $\times$ 4.2 & 0.1080 & 1 \nl
 3 & & 23 07 53.68 & -22 38 23.9 & & g & 1 & & 18.53 & 18.61  & & 9.5 $\times$ 4.4 & 0.1706 & 2 \nl
 4 & & 23 07 46.68 & -22 27 30.7 & & g & 1 & & 18.65 & 18.40  & & 8.2 $\times$ 6.8 & 0.1971 & 3,1\tablenotemark{\ddagger} \nl
 5 & & 23 07 25.43 & -22 27 26.6 & & g & 1 & & 18.69 & 18.65  & & 5.6 $\times$ 5.1 &  - & - \nl
 
 6 & & 23 07 19.23 & -22 52 52.0 & & g & 2 & & 19.67 & 18.56  & & 3.7 $\times$ 2.9 & 0.1689  & 4 \nl
 7 & & 23 08 31.98 & -22 42 30.7 & & g & 2 & & 19.68 & 18.42  & & 4.0 $\times$ 2.5 &  - & - \nl
 8 & & 23 07 26.27 & -22 25 01.5 & & g & 2 & & 19.71 & 18.42  & & 3.4 $\times$ 3.0 &  - &-  \nl
 9 & & 23 06 40.80 & -22 17 13.8 & & g & 2 & & 19.84 & 18.58  & & 3.5 $\times$ 3.0 & 0.1377  & 4 \nl
10 & & 23 06 37.02 & -22 43 47.5 & & g & 2 & & 19.96 & 18.62  & & 3.3 $\times$ 2.9 &  - & - \nl
 
11 & & 23 07 52.12 & -22 49 33.9 & & s & 2 & & 19.07 & 18.48  & & 3.4 $\times$ 2.8 &  - & - \nl
12 & & 23 07 22.41 & -22 15 20.9 & & s & 2 & & 19.19 & 18.56  & & 3.2 $\times$ 2.6 &  - & - \nl
13 & & 23 06 37.73 & -22 49 38.8 & & s & 2 & & 19.35 & 18.42  & & 3.4 $\times$ 3.0 &  - & - \nl
14 & & 23 06 52.30 & -22 49 56.9 & & s & 2 & & 19.48 & 18.53  & & 3.3 $\times$ 2.9 &  - & - \nl
15 & & 23 06 59.64 & -22 25 51.1 & & s & 2 & & 19.48 & 18.63  & & 3.7 $\times$ 2.6 &  - & - \nl
\enddata
\tablenotetext{\sharp}{ The codes are: g = 1 = galaxy; s = 2 = star.}
\tablenotetext{\ast}{ Major ($a$) and minor ($b$) axis in arcsecs.}
\tablenotetext{\dagger}{ Site of observation or reference; the codes are: 
(1)~CASLEO, Argentina; (2)~OPD, Brazil; (3)~Ciardullo, Ford \& Harms (1985); 
(4)~ESO, Chile.}
\tablenotetext{\ddagger}{ Mean velocity; difference of only $9 \ km~s^{-1}$ 
between the two measurements.}
\end{deluxetable}

\pagestyle{empty}
\makeatletter
\def\jnl@aj{AJ}
\ifx\revtex@jnl\jnl@aj\let\tablebreak=\nl\fi
\makeatother

\begin{deluxetable}{lrcrrcrr}
\small
\tablenum{6}
\tablecolumns{8}
\tablewidth{0pt}
\tablecaption{Results of Kolmogorov-Smirnov test for the color index 
 distributions\tablenotemark{\dagger}}
\tablehead{ \colhead{} & \colhead{} & \colhead{} & \multicolumn{2}{c}
 {$\times$ Galaxies in all (139)} & \colhead{} & \multicolumn{2}{c}
 {$\times$ Stars in all (177)} \nl
 \cline{4-5} \cline{7-8}
 \tablevspace{0.05cm}
 PDS galaxies: & Number & \colhead{} & $KS$ & $P_{KS}$ & \colhead{} &
 $KS$ & $P_{KS}$}
\startdata
Stars only in SSC         &  (91) & & 0.091 &  75.5 & & 0.158 &  10.1 \nl
Stars only in APM         &  (53) & & 0.082 &  95.8 & & 0.132 &  47.9 \nl
Stars both in SSC and APM &  (45) & & 0.225 &   6.4 & & 0.178 &  20.7 \nl
Galaxies in all           & (139) & & 0.029 & 100.0 & & 0.187 &   0.9 \nl
\enddata
\tablenotetext{\dagger}{ The numbers in parentheses refer to the size of
each sample.}
\end{deluxetable}

\pagestyle{empty}
\makeatletter
\def\jnl@aj{AJ}
\ifx\revtex@jnl\jnl@aj\let\tablebreak=\nl\fi
\makeatother

\begin{deluxetable}{cccccccccccc}
\small
\tablenum{7}
\tablecolumns{12}
\tablewidth{0pt}
\tablecaption{Examples of PDS area objects that are in the CCD images}
\tablehead{\colhead{} & \colhead{} & \multicolumn{2}{c} {Coordinates} 
& \colhead{} & \multicolumn{2}{c} {Classification\tablenotemark{\sharp}} 
& \colhead{} & \multicolumn{2}{c} {b$_J$ magnitude} & \colhead{} 
& \colhead{} \nl
 \cline{3-4} \cline{6-7} \cline{9-10}
\tablevspace{0.05cm}
Nr. & \colhead{} & $\alpha_{2000}$ & $\delta_{2000}$ & \colhead{} & PDS 
& SSC & \colhead{} & PDS & SSC & \colhead{}  
& $a \times b$ \tablenotemark{\ast} }
\startdata 
 1 & & 23 07 44.23 & -22 26 31.8 & & g & 2 & & 20.23 & 19.37  & & 3.3 $\times$ 2.3 \nl
 2 & & 23 07 43.51 & -22 33 32.5 & & g & 2 & & 20.37 & 19.72  & & 3.4 $\times$ 2.2 \nl
 3 & & 23 07 41.46 & -22 43 19.5 & & g & 2 & & 20.38 & 19.58  & & 2.8 $\times$ 2.5 \nl
 4 & & 23 07 48.31 & -22 28 16.1 & & g & 2 & & 20.39 & 19.82  & & 3.3 $\times$ 2.2 \nl
 5 & & 23 07 34.71 & -22 42 07.3 & & g & 2 & & 20.46 & 19.51  & & 3.9 $\times$ 2.1 \nl
\enddata
\tablenotetext{\sharp}{ The codes are: g = 1 = galaxy; s = 2 = star.}
\tablenotetext{\ast}{ Major ($a$) and minor ($b$) axis in arcsecs.}
\end{deluxetable}

\newpage
 
\figcaption[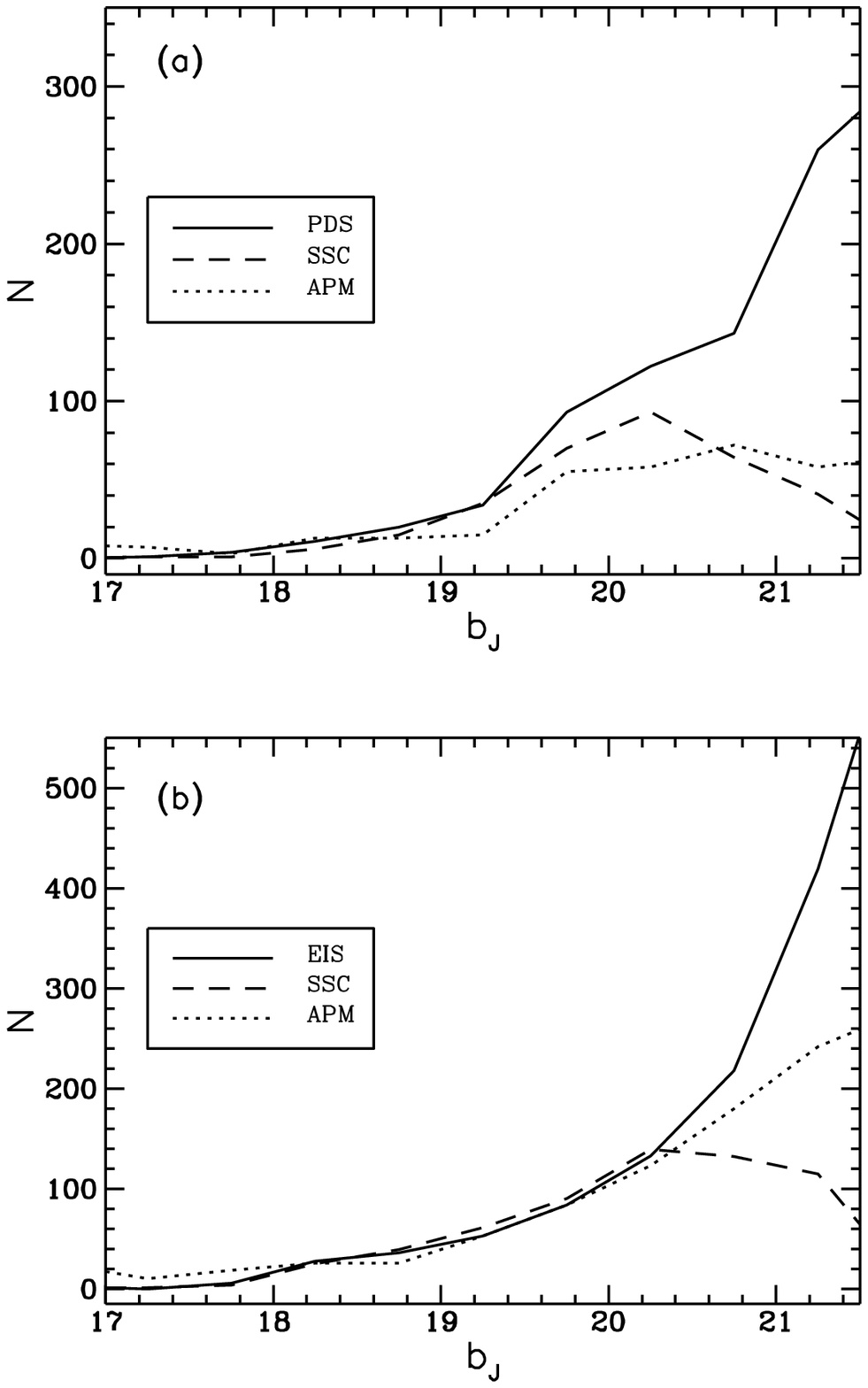]{($a$) Number counts for galaxies in the
3 catalogs of PDS area, and ($b$) number counts for galaxies in the 
3 catalogs of EIS area. \label{fig1}}
 
\figcaption[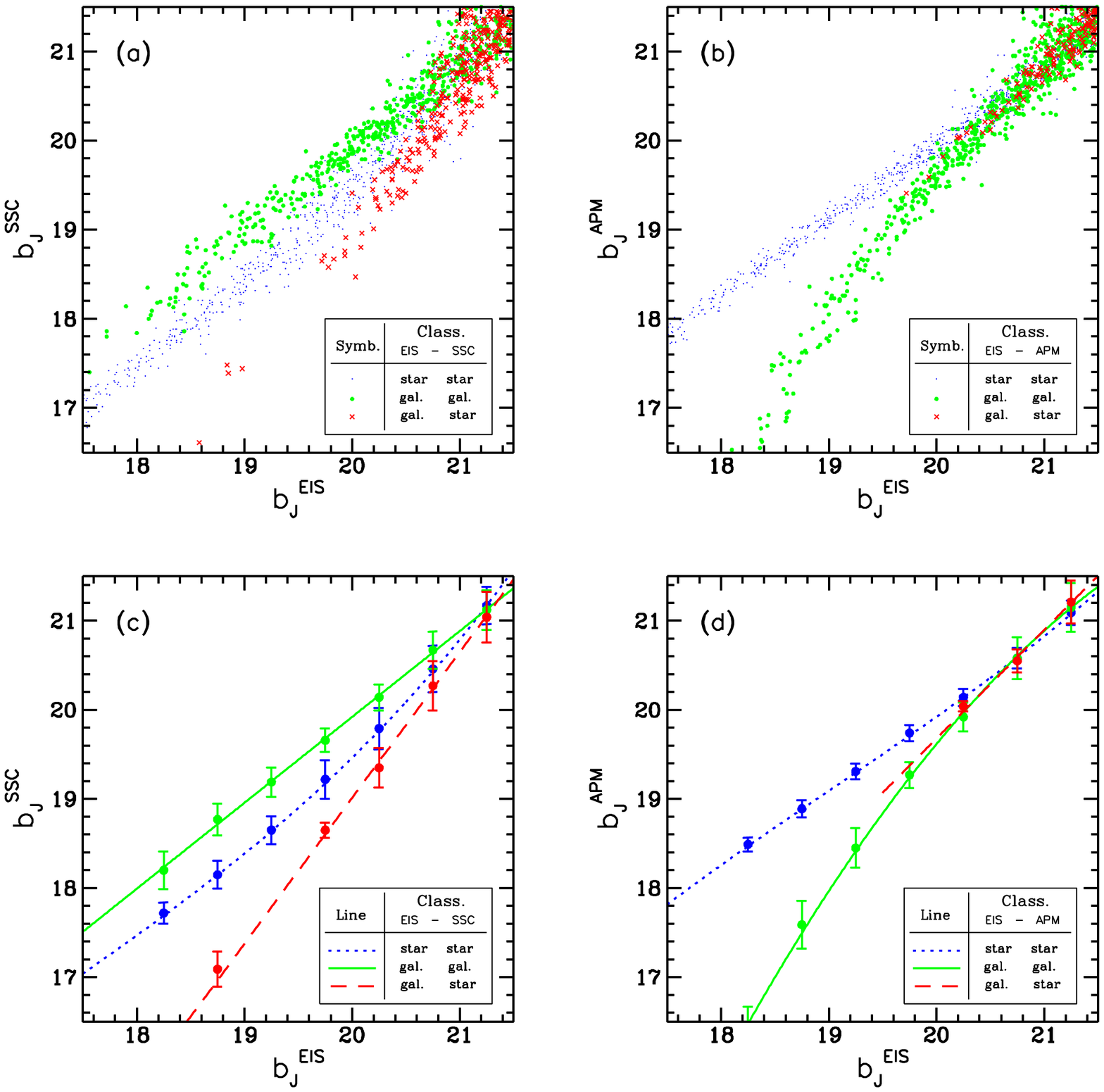]{($a$) Comparison between EIS and SSC 
magnitudes, and ($b$) between EIS and APM magnitudes.
The symbols denote the match in classification in each par of 
catalogs. ($c$) Polynomial fits to the three types of objects with
different symbols in panel ($a$), and ($d$) the same for the objects
in panel ($b$). The error bars represent the dispersion estimates of
each distribution. \label{fig2}}
 
\figcaption[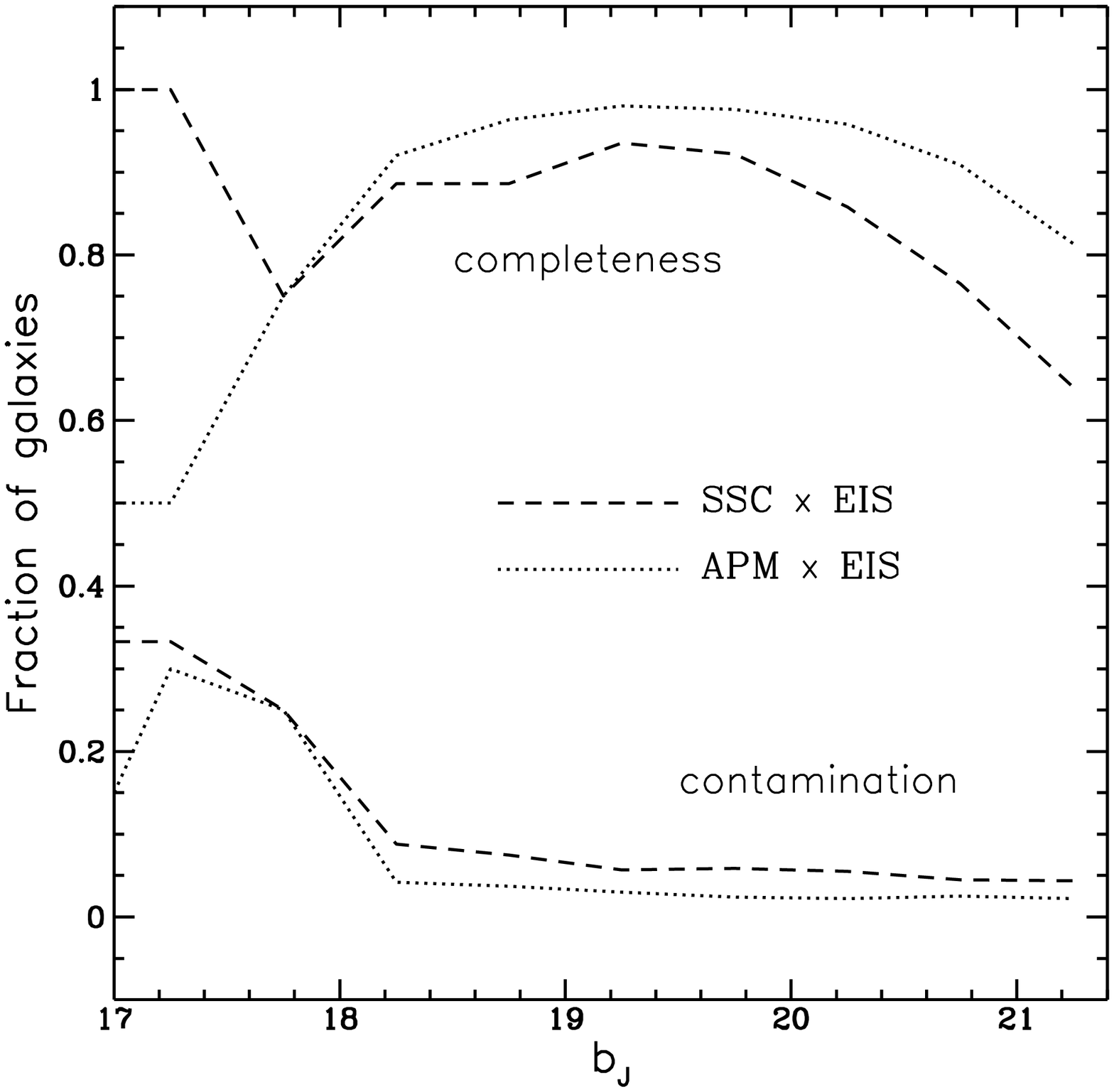]{Integrated completeness
and contamination levels for SSC and APM as compared to
EIS data. The features around $b_J =$~17.5-18.0 may be an artifact of 
small number statistics. \label{fig3}}
 
\figcaption[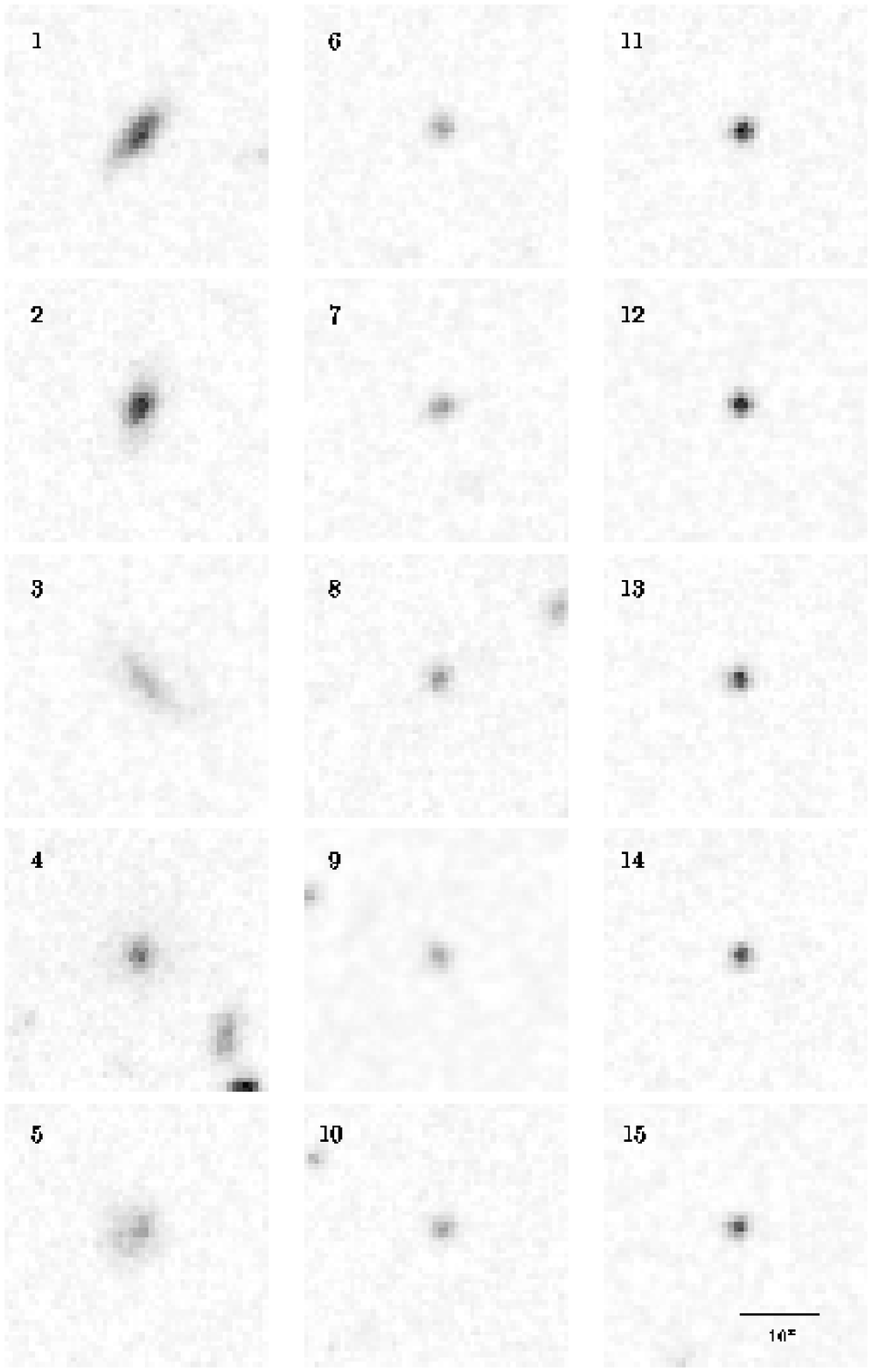]{Digitized images of the sample of objects
belonging to both PDS and SSC catalogs listed on Table 5.  
Objects classified as galaxies in both catalogs are displayed on the first 
column, objects classified as galaxies in PDS and stars in SSC on 
second column while objects classified as stars in both are displayed
in the third column. \label{fig4}}
 
\figcaption[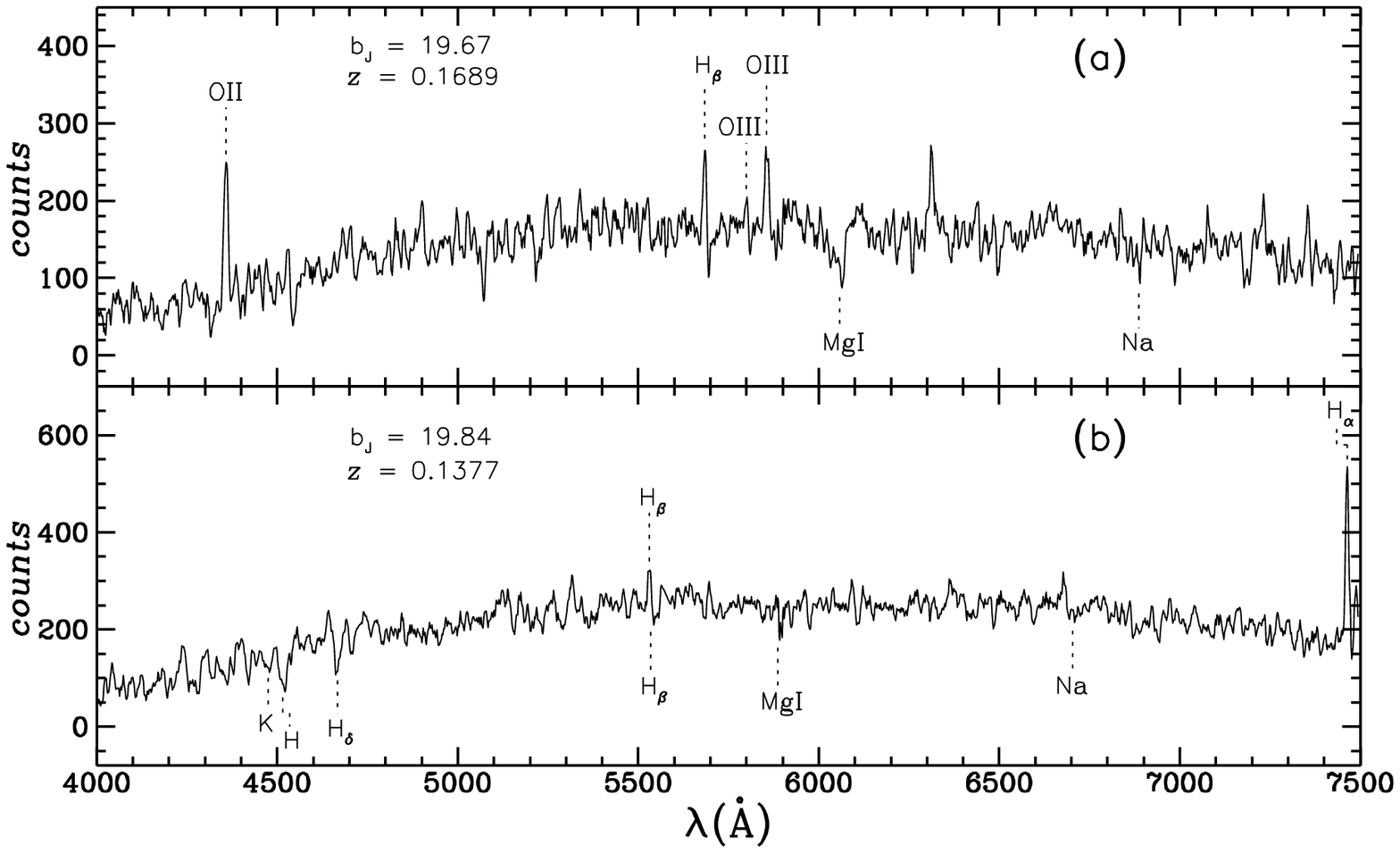]{Spectra obtained for two of the galaxies
misclassified in SSC, respectively numbers 6 (panel $a$) and 9 
(panel $b$) of Table 5. 
\label{fig5}} 
 
\figcaption[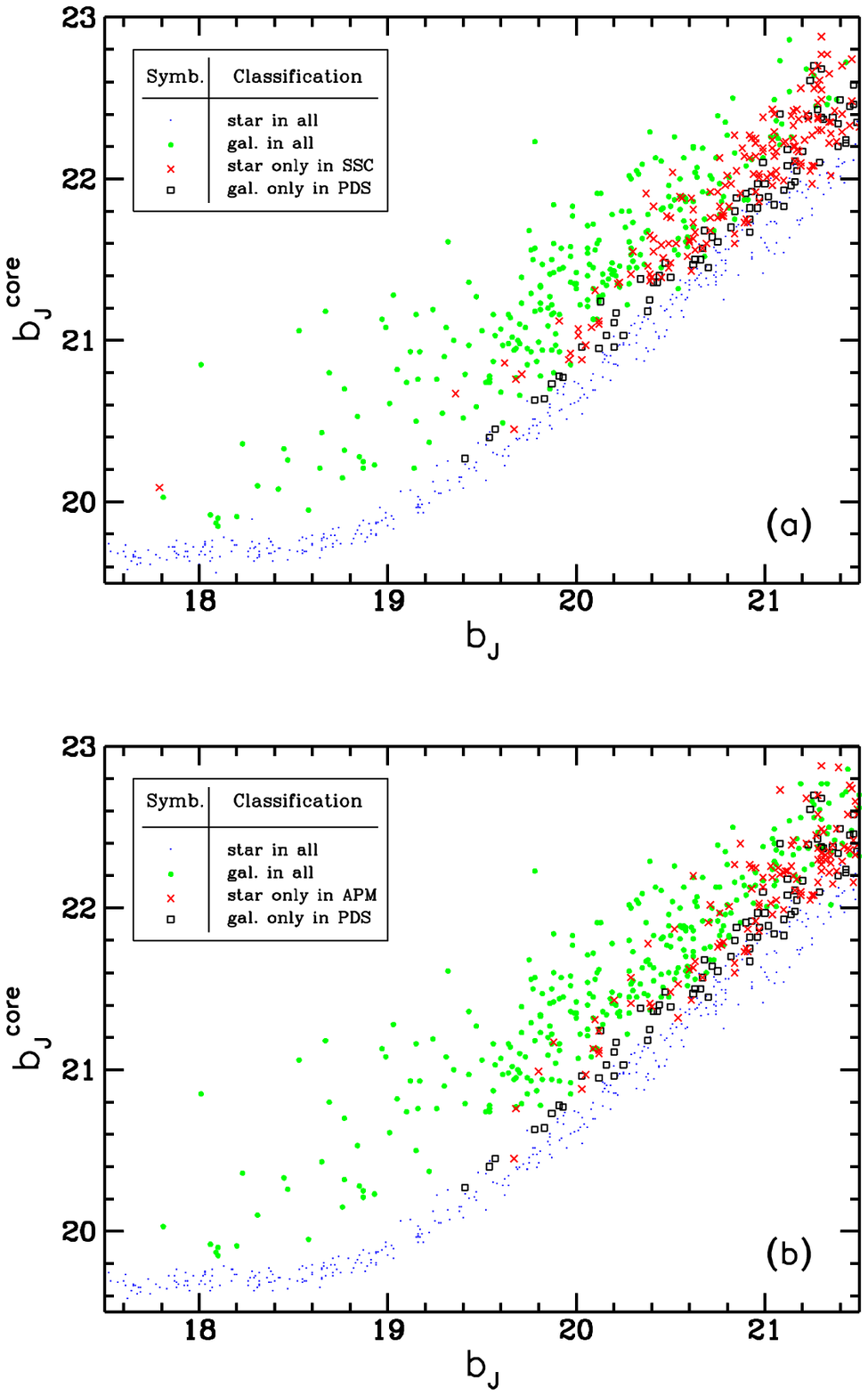]{Distribution of core magnitudes (relative
to the brightness of the 9 central pixels in the object's image) against
total magnitudes. Panel ($a$) represents the comparison between SSC and 
PDS, and panel ($b$) between APM and PDS. \label{fig6}}
 
\figcaption[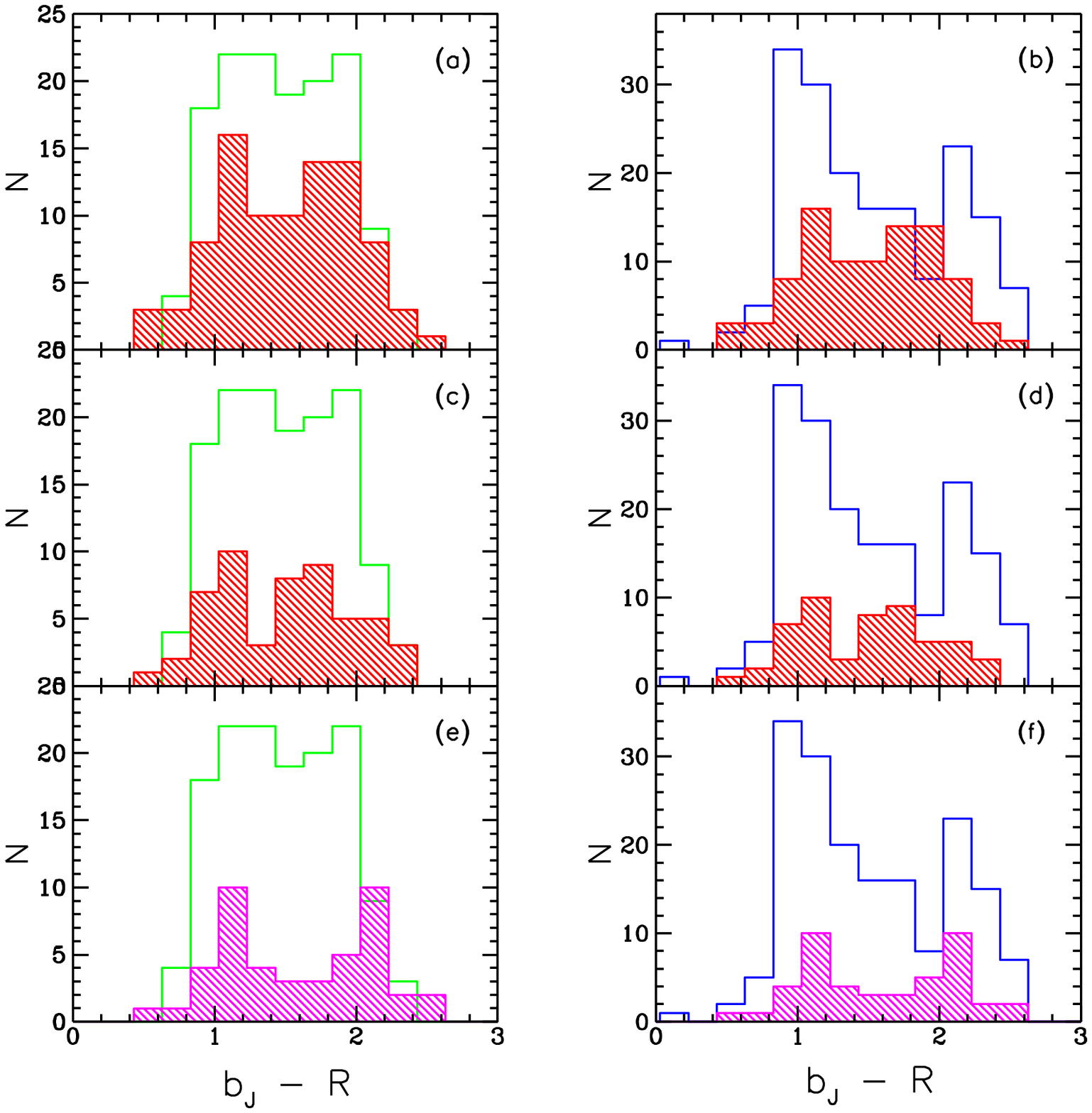]{Color index histograms for the 3 groups of
misclassified PDS galaxies (shaded histograms): stars only in
SSC (panels $a$ and $b$), stars only in APM (panels $c$ and $d$)
and stars in both (panels $e$ and $f$); compared to the (blank)
histograms of galaxies in all (first column) and stars in all
(second column). \label{fig7}}
 
\figcaption[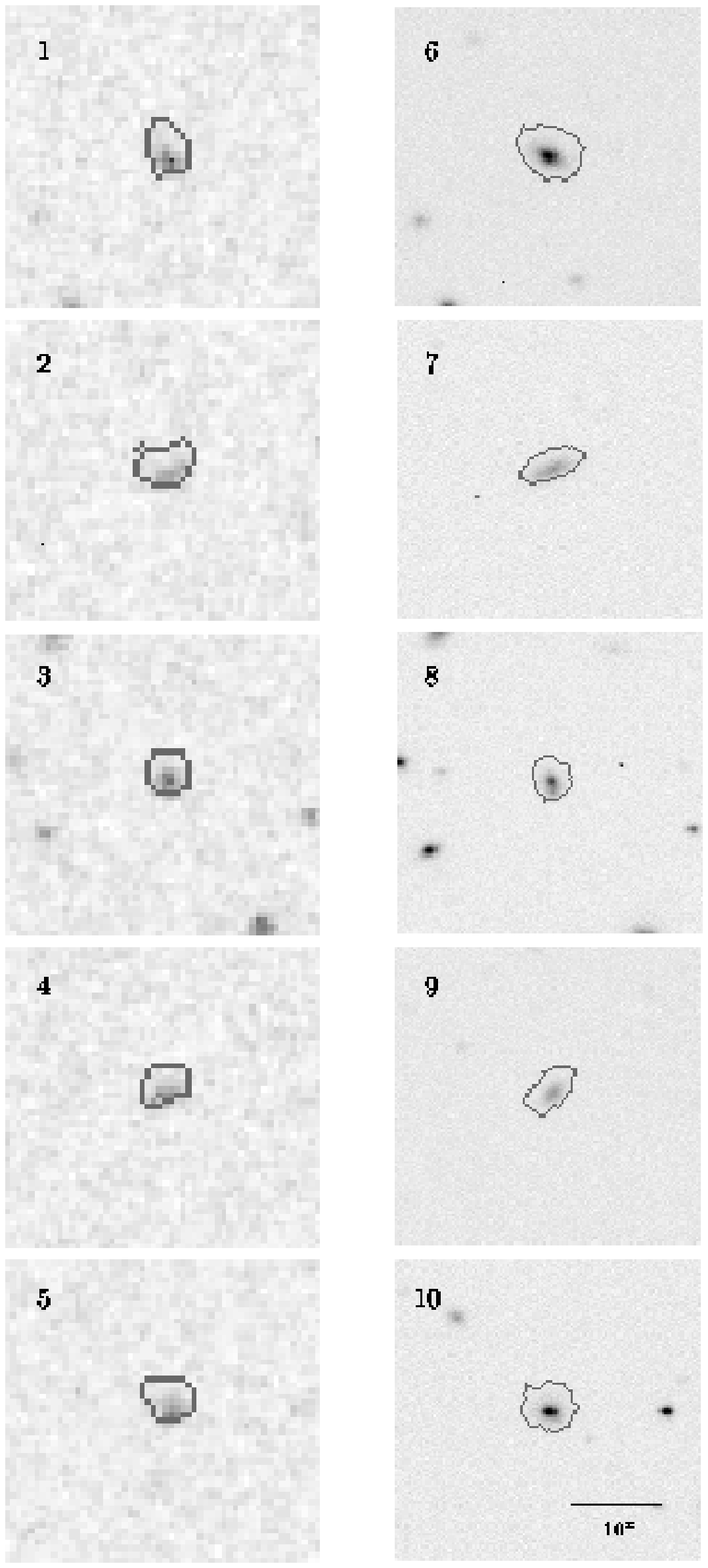]{PDS and CCD images for galaxies 
misclassified as stars only in SSC. The contours are FOCAS 
isophotes (displaced one pixel up). \label{fig8}}

\end{document}